\documentclass[preprint2]{aastex}

\def\deg{\hbox{$^\circ$}}

\shortauthors{H.M.Tovmassian, O.Yam, \& H.Tiersch}  
\shorttitle{X-ray emission of galaxy groups}

\begin{document}

\title{On the X-ray emission of the low-mass galaxy groups}

\author{H. M. Tovmassian and O. Yam}

\affil{Instituto Nacional de Astrof\'{\i}sica \'Optica y Electr\'onica,
AP 51 y 216, 72000, Puebla, Pue, M\'exico}   
\email{hrant@inaoep.mx, oyam@inaoep.mx}  

\and 
 
\author{H. Tiersch} 
 
\affil{Sternwarte K\"{o}nigsleiten, 81477, M\"{u}nchen, Leimbachstr.\ 1a 
Germany} 
\email{htiersch@uni.de}

\date{Received ...2001 / Accepted  .. 2001}

\begin{abstract} 

It is shown that the low-mass groups obey the $L_{x}\sim\sigma_v^{4}$ law
deduced for galaxy clusters. The impression of the more shallow slope of the
$L_x-\sigma_{v}$ correlation for groups is created not by enhanced X-ray
emission, but by underestimation of the radial velocity dispersion of some
groups.

\end{abstract}

\keywords{galaxies: clusters : general -- X-rays: galaxies}

\section{Introduction}

Solinger \& Tucker (1972) showed that if the source of the X-ray radiation is
hot gas bound in clusters, then the X-ray luminosity, $L_{x}$, should be
correlated with the radial velocity dispersion, $\sigma_v$. Thermal emission
from the intracluster gas yields an X-ray luminosity, $L_{x}$, proportional
to the square of the gas density. In the case of a constant mass-to-light
ratio, the $L_{x}$ is proportional to the square of the mass of the cluster.
If the cluster is a relaxed system, $\sigma_v$ is roughly proportional to the
square root of the mass. Thus, $L_{x} \propto \sigma_v^{4}$. Quintana \&
Melnick (1982) showed that $L_{x}$ of galaxy clusters, indeed, obeys the
expected correlation. Dell`Antonio et al. (1994) showed that rich groups
follow the $L_x \propto \sigma_{v}^4$ relation, but groups with smaller
$\sigma_v$ do have more shallow slope, $L_{x} \propto \sigma_{v}^{2.7}$. All
recent observations (Mahdavi et al. 1997, Zabludoff \& Mulchaey 1998, Helsdon
\& Ponman 2000, Mahdavi et al. 2000, Xue \& Wu 2000, Mahdavi et al. 1997)
proved that the dependence of the X-ray emission for low-mass groups (which
includes compact groups) on the $\sigma_v$ is much weaker than for galaxy
clusters. Zimer et al. (2001) analyzing the results of different investigators
mentioned some discrepancies, and concluded that they were due to poorly
determined $\sigma_v$s and $L_{x}$s. All data, show, however, that some amount
of low-mass groups of galaxies are located on the left side of the line $L_x
\propto \sigma_{v}^{\sim4}$. It has been generally assumed that the reason of
such location of groups on the graph $L_x - \sigma_{v}$ is the enhanced (by
one-two orders of magnitude) X-ray emission of groups. It is widely assumed
that the excess X-ray luminosity of the low-mass groups is explained by the
"mixed emission" scenario (Dell`Antonio et al. 1994) when the emission from
the intragroup plasma may be contaminated by a superposition of diffuse X-ray
sources corresponding to the hot interstellar medium of the member galaxies.
However, the shift of groups to the left of the $L_x \propto \sigma_{v}^{4}$
line may have another reason: underestimation of $\sigma_v$s.

It has been shown that HCGs and ShCGs have a triaxial spheroid, "cigar"-like
shapes (Malykh \& Orlov 1986, Hickson et al. 1984, Oleak et al. 1998). In a
series of papers Tovmassian and collaborators (Tovmassian et al. 1999,
Tovmassian \& Chavushyan 2000, Tovmassian et al. 2001) showed that
$\sigma_v$s of compact groups (CGs), and of associated with them loose groups
(LGs) which are also elongated and have the same orientation as corresponding
CGs, are correlated with elongation of groups determined by $b/a$
ratio\footnote{$a$ is the angular distance between the most widely separated
galaxies in the group, and $b$ is the sum of the angular distances $b_{1}$ and
$b_{2}$ of the most distant galaxies on either side of the line $a$ joining
the most separated galaxies (Rood 1979).}. It means that members of CGs and
LGs move along the elongation of corresponding group. It has been shown by
Tovmassian (2001a, 2001b) and Tovmassian \& Tiersch (2001) that out of three
possibilities of such movement: flying out of galaxies from the center of the
group in opposite directions, infalling from opposite directions, and regular
rotation of member galaxies in elongated orbits around the gravitational
center of each system, the latter possibility is the more realistic one. The
rotation time is less than $\sim3\times10^9$ years, so CG+LG systems may well
be virialized. The measured $\sigma_{v}$s of such elongated groups depend on
the orientation of the group. The highest values of $\sigma_{v}$s are observed
in those groups orientation of elongation of which is close to the line of
sight. Such groups generally have the highest $b/a$ ratio, though the
chain-like groups oriented at small angles $\theta$ to the line of sight
would also have relatively high $\sigma_{v}$. The measured $\sigma_{v}$s of the
majority of groups oriented at intermedient angles to the line of sight are
smaller. The groups oriented close to the orthogonal to the line of sight have
the smallest measured $\sigma_{v}$s, i.e. their $\sigma_{v}$s are highly
underestimated. The latter groups are the most elongated and have the smallest
$b/a$ ratio. Hence, they would be located on the left part of the
$L_{x}-\sigma_{v}$ graph. We show in this paper that underestimation of
$\sigma_{v}$s of the seen edge-on elongated groups, indeed, creates the more
shallow slope of the $L_{x}-\sigma_{v}$ correlation.

\section{Analysis of data}

For this analysis we used the dataset of Mahdavi et al. (2000) which is the
largest sample of the low-mass galaxy groups (RASSCALS) with detected X-ray
emission, and the X-ray selected sample of HCGs (Ponman et al. 1996). We
compared $b/a$ ratios of the groups most remote to the left from the line $L_x
\propto \sigma_{v}^4$ on the $L_x - \sigma_{v}$ graph with those of the most
remote to the right, or with those of the all other groups. The ratios $b/a$
for RASSCALS were determined on the corresponding maps presented in Mahdavi et
al. (2000), and those of for HCGs were taken from Tovmassian et al. (1998). In
the latter paper it is shown that the correlation between $\sigma_{v}$s and
$b/a$s does not depend on the number of group members. 

The following ten RASSCALS: SRGb063 (0.35), SRGb075 (0.60), SRGb119
(0.44), SSb085 (0.43), NRGb045 (0.61), SS2b153 (0.52), NRGs317 (0.43),
SRGb009 (0.27), SS2b293 (0.41), and SS2b313 (0.37) are the most remote to the
left (Fig. 1). The groups SS2b056 (0.87), NRGb004 (0.77), NRGs027 (0.62),
NRGb0032 (0.56), NRGs110 (0.50), NRGs117 (0.75), NRGb155 (0.96), NRGs388
(0.45), SS2b261 (0.46), and SRGs040 (0.87) are the most remote to the right.
The corresponding $b/a$ values are listed in parentheses. The other 49 groups
with measured X-ray emission are located along the $L_x \propto
\sigma_{v}^4$ line within the stripe with $log \sigma_v=\pm0.1$ width. The
mean $b/a$ of the groups at the left is equal to $0.44\pm0.11$, and that of the
groups at the right is $0.68\pm0.19$.

The same trend is observed in the case of HCGs. The mean value of $b/a$s of
four groups located most remote to the left of the line $L_x-\sigma_{v}^4$
(Fig. 2) is equal to $0.20\pm0.05$, while that of the rest 18 groups is equal
to $0.44\pm0.18$. The Kolmogorov-Smirnov test rejected the hyphotesis that
the compared distributions are of the same parent distribution in both
considered cases. The significance level of rejection for the RASSCALS is
$\alpha=0.0069$, and for HCGs is $\alpha=0.0082$.

\begin{figure}[htb]
\includegraphics[angle=90,width=7cm,keepaspectratio]{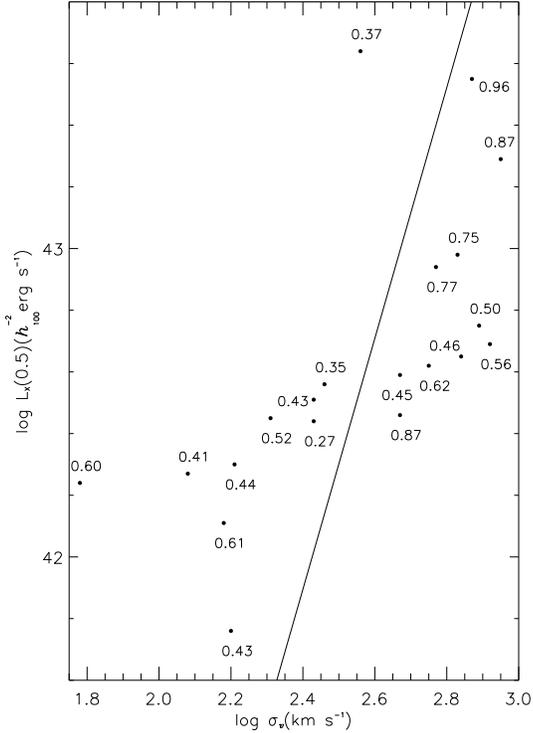}
\figcaption[fig1.eps]{$L_x - \sigma_v$ relation of groups. Data are taken from
Mahdavi et al. (2000). Only the groups located at the utmost left and the
utmost right from the line $L_x \propto \sigma_{v}^4$ are drawn. The $b/a$
ratios are shown.} \label{fig1}
\end{figure}

\begin{figure}[htb]
\includegraphics[angle=90,width=7cm,keepaspectratio]{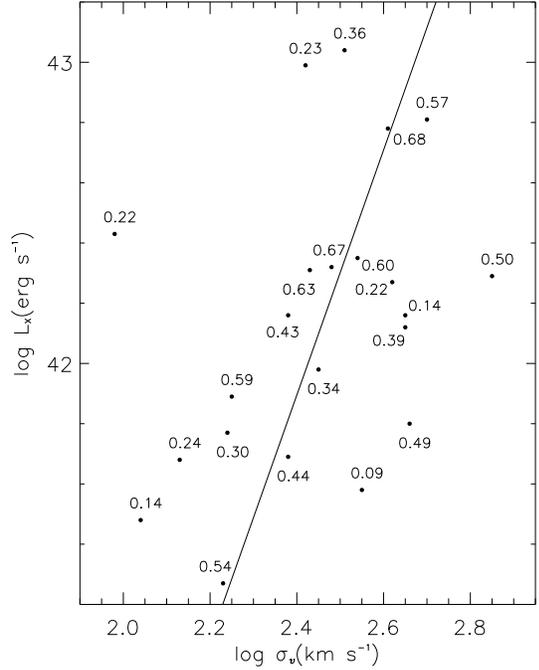}
\figcaption[fig2.eps]{$L_x - \sigma_v$ relation for HCGs (from Ponman et
al. 1996). The line $L_x \propto \sigma_{v}^4$ is drawn as in Fig. 1. The $b/a$
ratios are shown.}
\label{fig2}
\end{figure}

Hence, the groups located to the left of the $L_x \propto \sigma_{v}^4$ line
on the $L_x - \sigma_{v}$ graph have the smallest $b/a$ ratios. It indicates,
that they are oriented close to the orthogonal to the line of sight. For this
reason their measured radial velocity dispersions are smaller than the real
values. Location of such groups on the left side of the $L_x \propto
\sigma_{v}^{4}$ line creates an impression of the more shallow slope of the
$L_x - \sigma_{v}$ correlation. Possible errors in $b/a$ values caused by
missed members or chance interlopers would, apparently, increase the
dispersion of the mean values, but would not introduce systematic errors in
either direction.

The highest $\sigma_{v}$s of HCGs are $\sim600$ km s$^{-1}$. The highest
$\sigma_{v}$s of the groups studied by Mahdavi et al. (2000) are of the same
order. Apparently, these are the groups observed end-on. Their measured
$\sigma_{v}$s generally correspond to the real values. Such groups would be
located at the utmost right of the $L_x \propto \sigma_{v}$ graph. With the
increase of the angle $\theta$ the $b/a$ ratio and the corresponding
$\sigma_{v}$ of the group would decrease. At angles $\theta$ of about
$75\deg$ the measured $\sigma_{v}$s would be by about 400 km s$^{-1}$ smaller
in comparison to the real values, and $log \sigma_{v}$ would be about 2.2,
which corresponds to that of the groups located at the utmost left on the $L_x
- \sigma_{v}$ graph.   

Thus, {\it not} the X-ray luminosities of the low-mass groups are higher than
it is expected by the $L_x \propto \sigma_{v}^4$ law, but the $\sigma_{v}$s of
them are {\it underestimated}. Rich clusters have much more symmetric round
shape, and orientation may have negligible effect on the measured $\sigma_{v}$.
The spread of clusters around the $L_x \propto \sigma_{v}^{4}$ line is due to
the natural dispersion of parameters.

\section{Conclusions}

Consideration of the $b/a$ ratios of the low-mass groups shows that groups
located at the utmost left of the $L_x \propto \sigma_{v}^4$ line on the $L_x -
\sigma_{v}$ graph have, on average, smaller $b/a$ values than those located at
the utmost right. The elongation of groups with small $b/a$ ratios are
oriented close to the orthogonal to the line of sight. The measured
$\sigma_{v}$s of these groups are, thus, underestimated, and are smaller than
the real values. Therefore, such groups are artificially shifted to the left
on the $L_x - \sigma_{v}$ graph. This creates an impression of an enhanced
X-ray luminosity. If to take into account the reasonable amount of
underestimation of $\sigma_{v}$s (of the order of 200-300 km s$^{-1}$, which
corresponds to angle $\theta$ of about $40\deg-50\deg$), then the corresponding
groups will be moved to the right, towards the line $L_x \propto \sigma_{v}^4$.
Hence, the low-mass groups {\it obey} the $L_x \propto \sigma_{v}^4$ law for
clusters of galaxies (Solinger \& Tucker 1972, Quintana \& Melnick 1982). It
means that there is no need to apply any mechanism of the enhancement of the
X-ray luminosity of the low-mass groups, since in reality there is no any
enhancement. 

Compact groups are stable systems with members probably rotating around the
gravitational center of the corresponding group (Tovmassian 2001a, 2001b).
For the reason of regular movement in elongated orbits the velocities of
member galaxies in the central region of a CG are high enough. Therefore, the
efficiency of interaction in such groups would be smaller than in the made
numerical simulations when such regular movement has been neglected (Barnes
1985, 1989; Ishizawa 1986; Mamon 1987, 1990; Zheng et al. 1993). The formation
of the hot interstellar medium in member galaxies, widely assumed for
explanation of the excess X-ray emission, may be very rare, and may not
dominate the global X-ray emission. In fact, the claimed by many excess of the
X-ray luminosity in the low-mass groups is due to a projection effect, and is,
thus, a result of {\it misinterpretation} of the observational data. The
finding of Mahdavi \& Geller (2001) that clusters of galaxies and single
elliptical galaxies form a continuous relation $L_x - \sigma_{v}^m$ are
consistent with the result presented in this paper. Low-mass groups are not
exotic objects and obey the same law. 

\begin{acknowledgements} We are grateful to the first Referee of our paper, Dr.
A.Mahdavi, and especially to the anonymous Referee for very valuable comments
which allowed us to improve the paper.
\end{acknowledgements}

\end{document}